\documentclass[a4paper,conference]{IEEEtran}
\IEEEoverridecommandlockouts
\usepackage{cite}
\usepackage{amsmath,amssymb,amsfonts}
\usepackage{algorithmic}
\usepackage{graphicx}
\usepackage{textcomp}
\usepackage{xcolor}
\usepackage{booktabs}
\usepackage{array}
\usepackage{url}
\usepackage{ulem}
\usepackage{hyperref}
\def\BibTeX{{\rm B\kern-.05em{\sc i\kern-.025em b}\kern-.08em
    T\kern-.1667em\lower.7ex\hbox{E}\kern-.125emX}}
    
\begin{document}

\title{UrbanEMF: City-Scale EMF Mapping over a Continuous Urban Area with Real-World Base-Station Deployment
}

\author{
\IEEEauthorblockN{
Shuangning Li\textsuperscript{$*$},
Chenxin Luo\textsuperscript{$*$},
Shanshan Wang\textsuperscript{$*$},
Yarui Zhang\textsuperscript{$\dagger$},
Paul Lagouanelle\textsuperscript{$*$},
Joe Wiart\textsuperscript{$*$}
}

\IEEEauthorblockA{
\textsuperscript{$*$}LTCI, Télécom Paris, Institut Polytechnique de Paris, Palaiseau, France\\
\textsuperscript{$\dagger$}SATIE, École normale supérieure Paris-Saclay, CNRS, Université Paris-Saclay, Gif-sur-Yvette, France\\
Corresponding Author: Shuangning Li. E-mail: shuangning.li@telecom-paris.fr
}
}

\maketitle

\begin{abstract}

Electromagnetic Fields (EMF) mapping is essential for wireless propagation modeling and are fundamental to a wide range of applications, including spectrum awareness, network planning, and integrated sensing and communication (ISAC). However, existing datasets are often limited to spatially separated 2D scenes without real base station (BS) information. To address this gap, we present UrbanEMF, a city-scale EMF mapping dataset constructed from real world urban geometry and physical BS deployments using ray tracing. Unlike conventional  single scene-based datasets,
UrbanEMF preserves continuous city-scale urban coverage and realistic transmitter location information. It further extends traditional 2D maps by incorporating multiple receiver heights. In addition, both aggregated received signal strength (RSS) maps and path loss maps are provided to characterize complementary aspects of the radio environment. This work provides a realistic and flexible benchmark for developing and evaluating learning-based methods in large-scale urban wireless environments. The code for this work is available at: \url{https://github.com/lemonstudy/UrbanEMF}

\end{abstract}

\begin{IEEEkeywords}
6G, EMF mapping, received signal strength (RSS), path loss
\end{IEEEkeywords}

\section{Introduction}

Future sixth-generation (6G) wireless networks are expected to deliver substantially higher data rates, ultra-low latency, massive connectivity, enhanced reliability, and native sensing and intelligence capabilities \cite{itu_m2160}. 
Realizing these capabilities increasingly requires networks to acquire prior knowledge of location-dependent propagation conditions, since functions such as beamforming, resource allocation, localization, and sensing can benefit from an accurate characterization of signal behavior across the wireless environment. Radio maps address this need by representing channel quantities such as received power, path loss, or interference as continuous functions of space, and thus offer a concrete foundation for environment-aware operation in 6G systems \cite{feng2025recent}.

From a broader perspective, EMF mapping is closely related to radio map construction, as both aim to represent communication signal conditions across different frequency bands. Although radio maps can describe a variety of radio frequency (RF) quantities, two types have been particularly widely studied \cite{feng2025recent}. The first type focuses on received signal characteristics, such as exposure maps derived from the accumulated received signal strength (RSS) contributed by multiple transmitters. The other describes propagation characteristics, with path loss maps quantifying the attenuation between a transmitter and receiver locations. 

Recent studies have shown promising results at using deep learning approaches to build accurate radio map in urban environments \cite{zhang2026exposnet,li2026phy2}. The use of extensive datasets is crucial to insure the accuracy of such models. Thus, a growing number of public datasets have been recently proposed. Notably, for path loss maps, RadioMapSeer provides a large collection of ray tracing urban maps and has become a widely adopted benchmark for radio map construction \cite{levie2021radiounet}. Other datasets extend this setting to different environments, antenna configurations, or three-dimensional representations. More recently, SpectrumNet introduced multiband 3D radio maps spanning multiple terrain and climate scenarios, significantly increasing the diversity of available propagation data \cite{zhang2025spectrumnet}. Related works such as RadioGAT further demonstrate the value of structured radio map datasets for multi-band reconstruction tasks \cite{li2024radiogat}. The survey in \cite{feng2025recent} also summarizes representative datasets such as RMDirectional Berlin \cite{jaensch2024radio} and CKMImageNet \cite{wu2025ckmimagenet}. Nevertheless, most of these datasets are organized as collections of individual simulation scenes or predefined transmitter configurations. Consequently, they do not fully preserve the spatial continuity of a complete real city together with its physical operational BS deployment.

In contrast, measurement-based datasets are assembled from physical measurements, which grounds them in reality but limits their spatial resolution. In France, the National Frequency Agency (ANFR) provides nationwide simulated RF-EMF exposure maps through the Cartoradio platform \cite{anfr_cartoradio}. These maps are generated from geographical information and authorized mobile-network station data and are intended for large-scale public exposure assessment. In parallel, DoNext contains extensive 4G and 5G measurements collected over approximately two years in Dortmund and includes both passive network observations and active measurements for spatial network analysis \cite{schippers2025donext}. Other measurement-based studies have also shown that RF-EMF levels can vary substantially with microenvironment, deployment conditions, and network configuration \cite{wang2025comprehensive,manassas2025distance}. However, measurement campaigns are inherently constrained by acquisition routes, measurement density, observation time, and collection cost. Similarly, public exposure maps are primarily designed for exposure monitoring rather than as configurable research datasets providing both aggregated radio conditions and transmitter-specific propagation information.

To address the limitations of current datasets, we propose \textit{UrbanEMF}, a flexible city-scale EMF map dataset generated through a configurable Sionna-based pipeline. The main contributions of this paper are summarized as follows:

\begin{enumerate}

    \item UrbanEMF is introduced as a large-scale dataset constructed from a city-scale model of the Paris urban environment. The framework can also be extended to other cities or regions by replacing the urban environment and BS deployment data.

    \item The cellular network is incorporated by using the exact locations of BSs instead of randomly generated transmitter positions. Antenna tilt is further considered in the propagation modeling to represent the directional characteristics of practical BS deployments. These settings combined allows us to simulate a realistic propagation study.
    
    \item The dataset provides two complementary radio maps displaying the accumulated RSS and the path loss. These features support height-aware radio environment analysis as well as different application tasks, ranging from coverage and exposure assessment to link-level propagation studies.
\end{enumerate}

\section{The UrbanEMF Dataset}

\subsection{Problem Formulation}
Consider a geographical region covered by a set of $K$ BSs. The region is uniformly divided into an $H \times W$ grid, where each grid cell represents a square area with side length $\Delta$. The physical environment and the BS deployment are represented as a multi-channel tensor of size $C \times H \times W$. The environmental channel $C$ describes spatial properties that affect radio propagation, such as obstacle occupancy and building geometry. The accumulated RSS at grid location $\mathbf{x}_{i,j}$ is denoted by:

\begin{equation}
P_{r}(\mathbf{x}_{i,j}),
\label{eq:received_power}
\end{equation}

where $P_{r}$ is expressed in dBm. In this work, the RSS values are generated using the Sionna ray-tracing simulator, which first computes the total received power from all BSs nearby in watts and then converts it into dBm according to:

\begin{equation}
P_{r}(\mathrm{dBm})
=
10\log_{10}\!\left(P_{r}(\mathrm{W})\right)+30.
\label{eq:watt_to_dbm}
\end{equation}

In Sionna, the ray-tracing simulation accounts for physical propagation mechanisms including line-of-sight propagation, specular reflection, and diffraction caused by the surrounding environment \cite{sionna}.

For each BS, we additionally provide a path loss map to characterize the \sout{large-scale} attenuation between the transmitter and every receiver location. So the path loss associated with the $k$-th BS at $\mathbf{x}_{i,j}$ is expressed in dB as:

\begin{equation}
L_{k}(\mathbf{x}_{i,j})
=
P_{t,k}-P_{r,k}(\mathbf{x}_{i,j}),
\label{eq:path_loss}
\end{equation}

where $P_{t,k}$ denotes the transmit power of the $k$th BS in dBm. The received power $P_{r,k}(\mathbf{x}_{i,j})$ is obtained directly from the propagation paths, and therefore the resulting path loss map inherently reflects the attenuation caused by distance.

\subsection{Key Characteristics}
\begin{table*}[htbp]
\centering
\caption{Comparison with Existing Radio Map and Wireless Channel Datasets}
\label{tab:dataset_comparison}
\small
\setlength{\tabcolsep}{4.5pt}
\begin{tabular}{@{}lccccccc@{}}
\toprule
\textbf{Dataset} &
\textbf{Simulation area} &
\textbf{Resolution} &
\textbf{Real BS} &
\textbf{Multi-Tx} &
\textbf{Multi-Height} &
\textbf{RSS} &
\textbf{Path Loss} \\
\midrule
\textbf{UrbanEMF (Ours)}
& $18.0 \times 9.6\,\mathrm{km}^2$ & $flexible$ & $\checkmark$ & $\checkmark$
& $\checkmark$ & $\checkmark$ & $\checkmark$ \\

RadioMapSeer~\cite{levie2021radiounet}
& $256\times256$~m$^2$ & $1\,\mathrm{m}$ & $\times$ & $\times$
& $\times$ & $\times$ & $\checkmark$ \\

UrbanRadio3D~\cite{wang2025radiodiff}
& $256\times256$~m$^2$ & $0.5\mathrm{m}$ & $\times$ & $\times$
& $\checkmark$ & $\times$ & $\checkmark$ \\

RMDirectionalBerlin~\cite{jaensch2024radio}
& $256\times256$~m$^2$ & $1\,\mathrm{m}$ & $\times$ & $\times$
& $\times$ & $\times$ & $\checkmark$ \\

CKMImageNet~\cite{wu2025ckmimagenet}
& $700\times700$~m$^2$ & $2\,\mathrm{m}$ & $\times$ & $\checkmark$
& $\checkmark$ & $\times$ & $\checkmark$ \\

DeepMIMO~\cite{deepmimo}
& $600\times440$~m$^2$ & $0.1,0.2\,\mathrm{m}$ & $\times$ & $\checkmark$
& $\checkmark$ & $\times$ & $\checkmark$ \\
\bottomrule
\end{tabular}
\end{table*}

Table~\ref{tab:dataset_comparison} compares UrbanEMF with commonly used other radio map datasets. RadioMapSeer \cite{levie2021radiounet} provides high resolution 2D path loss maps over fixed size urban scenes, where each data sample is associated with simulated placed transmitters. UrbanRadio3D \cite{wang2025radiodiff} extends this local scene setting by incorporating three-dimensional urban geometry and receiver-height information, but the transmitter deployment remains artificially configured. RMDirectionalBerlin \cite{jaensch2024radio} is constructed from real locations in Berlin and includes realistic building heights, vegetation, and directional rooftop antennas. However, each radio map is still generated independently for a selected transmitter rather than representing an operational multi-BS deployment. CKMImageNet \cite{wu2025ckmimagenet} focuses on channel knowledge maps across predefined urban, rural, and indoor scenarios, providing rich channel information such as channel gain, AoA, AoD, and propagation delay together with environmental representations. In contrast, DeepMIMO \cite{deepmimo} is a parameterized wireless channel dataset generated from predefined ray-tracing scenarios and dense Tx--Rx links, targeting mainly mmWave and massive MIMO applications rather than city-scale radio mapping.
Hence, the key characteristics of UrbanEMF are further detailed below.

\textbf{Real World BS Deployment.}
UrbanEMF uses the geographical locations of operational BSs recorded in Cartoradio, the official antenna database maintained by the ANFR \cite{anfr_cartoradio}, instead of relying on randomly generated transmitter positions. Therefore, the simulated radio maps preserve the realistic relationship between the actual urban geometry and the physical BS deployment.

\textbf{Continuous Urban Coverage.}
Instead of constructing independent urban scenes, UrbanEMF models the Paris urban area as a spatially continuous environment. The city is simulated district by district with additional surrounding areas. This design better preserves large-scale spatial continuity than datasets composed of mutually independent scenes, which differs from the commonly adopted single-transmitter setting and provides radio environments that are closer to practical dense urban networks.

\textbf{Multi-Height Propagation Information.}
UrbanEMF extends conventional ground-level 2D EMF maps by evaluating the radio environment at multiple receiver heights. For the same horizontal location, propagation results can therefore be obtained at different vertical positions. These height-dependent maps provide additional information for studying three-dimensional and height-aware radio propagation in urban areas.

\textbf{Flexible EMF Map Outputs.}
Our dataset provides both accumulated RSS and path loss maps. Accumulated RSS maps represent the spatial distribution of received signal power under a given transmission configuration and are therefore well suited for coverage evaluation, exposure assessment, and other area-based analyses. In contrast, path loss maps describe the propagation attenuation between a specific BS and receiver locations, making them more appropriate for link-level analysis, channel characterization, and point-to-point propagation studies. By providing both representations, UrbanEMF can support a broader range of radio-environment research tasks.

\section{Generation Process}

The dataset generation process includes real-world data acquisition, urban-scene construction and Sionna-based ray-tracing simulation. The remainder of this section discusses this process in detail and shows the simulation results.

\subsection{Data Acquisition and Scene Construction}

The geographical environment is derived from \href{https://www.openstreetmap.org/}{OpenStreetMap} (OSM), while the BS deployment is obtained from Cartoradio, the official antenna database maintained by the (ANFR). After filtering the records antennas at $2.6$~GHz, a total of $1,992$ operational BSs located within the 20 administrative districts of Paris are conserved. According to statistics, the transmitter heights are mainly concentrated around several tens of meters, with a median value of approximately $28.3$~m.

The OSM building data are processed district by district and transformed from geographic coordinates into the Lambert-93 projected coordinate system (EPSG:2154), which is used throughout the simulation. Each district is extended by a $50$~m surrounding area so that propagation near administrative boundaries can account for neighboring geometry. The resulting vector building footprints are converted into three-dimensional meshes and exported as Mitsuba XML scenes compatible with Sionna RT.

\begin{table}[htbp]
\centering
\caption{Main Simulation Configuration}
\label{tab:radio_config}
\small
\setlength{\tabcolsep}{4pt}
\begin{tabular}{@{}ll@{}}
\toprule
\textbf{Parameter} & \textbf{Value} \\
\midrule
Region of Interest & Paris, 20 districts \\
BS database & Cartoradio / ANFR \\
Number of BSs & 1,992 \\
Carrier frequency & $2.6$~GHz \\
Simulation coordinates & Lambert-93 (EPSG:2154) \\
District buffer & $50$~m \\
Tx power & $40$~dBm \\
Tx antenna & $1\times1$ dipole, cross-polarized \\
Rx antenna & $1\times1$ isotropic, V-polarized \\
Building material & ITU concrete \\
Ground material & ITU medium dry ground \\
Reference RSS Rx height & $1.5$~m \\
Maximum interaction depth & 16 \\
\bottomrule
\end{tabular}
\end{table}

Building heights are determined from OSM attributes, using explicit height values when available and otherwise estimating them from the reported number of building levels with an uncertainty of 3 meters. When neither source is available, a default height of $15$~m is adopted. Building and ground surfaces are assigned ITU radio materials corresponding to concrete and medium dry ground, respectively. A flat ground model is used because terrain elevation variations within central Paris are relatively limited compared to the dominant building geometry.

The transmitter height is directly taken from the Cartoradio ``height above ground'' record. Therefore, the simulated vertical coordinate is defined relative to the local ground level rather than by adding the rooftop height. The transmitter array is modeled as a $1\times1$ dipole antenna with cross polarization, while the receiver uses a $1\times1$ isotropic vertically polarized antenna.

To reduce peak memory usage during large-area ray-tracing simulations for RSS map generation, the transmitters are divided into multiple batches for processing, while the receiver grid is divided into bounded subregions. Each subregion still accounts for all relevant transmitters through sequential batch computation, and the resulting received powers are incrementally aggregated before the subregions are stitched into the complete district-level map.

\begin{figure*}[htbp]
    \centering
    \includegraphics[width=\textwidth]{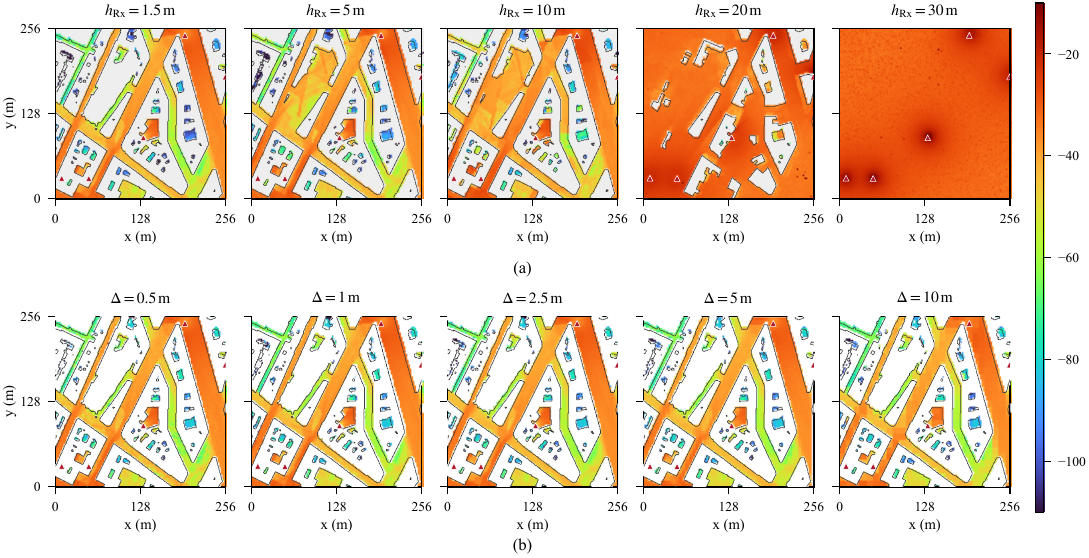}
    
    \caption{\fontsize{9pt}{10pt}\selectfont
    Example accumulated RSS maps of the same urban region at different receiver heights. The spatial distribution gradually changes from strongly building-dependent propagation at lower heights to more uniform coverage above the surrounding urban structures.}
    \label{fig:rss_xy}
\end{figure*}

\subsection{Sionna-Based EMF Map Simulation}

This section describes the generation of the two complementary EMF map representations, namely exposure and path loss maps.

\subsubsection{Accumulated RSS Map}

Exposure maps are generated using Sionna's RadioMapSolver. For each district, all relevant BSs in the simulation scene are jointly considered and the received power is evaluated over a regular horizontal grid with $1$~m resolution.  Thus, the accumulated RSS value at each location represents the aggregate contribution of all BSs within the simulated scene.

Fig.~\ref{fig:rss_xy} illustrates this height-dependent behavior for the same local urban region. At low receiver heights, the accumulated RSS distribution is strongly shaped by streets and surrounding buildings, and clear shadowed regions are visible behind building blocks. As the receiver height increases, these local obstructions become progressively less dominant and the spatial distribution becomes smoother. At sufficiently high receiver positions, the influence of rooftop blockage is substantially reduced and the received power is mainly governed by the large-scale transmitter geometry.

\subsubsection{Path Loss Map}

\begin{figure}[htbp]
    \centering
    \includegraphics[width=\columnwidth]{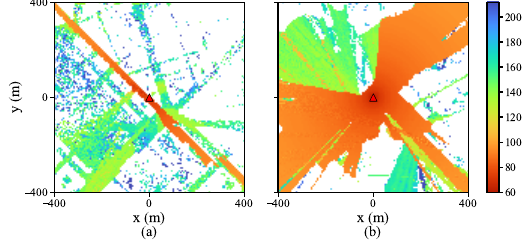}
    \caption{Representative path loss maps under different urban environments: (a) a relatively dense urban scene and (b) a more open urban scene.}
    \label{fig:path_loss}
\end{figure}

In addition to accumulated RSS maps, an independent single BS path loss pipeline is implemented using Sionna's PathSolver. For each transmitter, an $800\times800$~m$^2$ receiver region centered on the BS is constructed with a spatial resolution of $5$~m. Since the path loss map focuses on the signal fading between a transmitter and receiver link rather than the total signal strength from all transmitters at every receiver, the area covered by path loss here doesn't take into account all the transmitters across the entire Paris area like an RSS map does. When a BS is located near a district boundary, the simulation region extends into neighboring districts to preserve the surrounding urban environment and avoid truncation at district borders.

For each Tx--Rx link, PathSolver resolves the multipath propagation components and returns the corresponding complex channel coefficients. The link gain is then computed from the sum of the resolved path coefficients.

Figure~\ref{fig:path_loss} shows representative path loss maps generated for two different urban environments. The results show that the path loss distribution is strongly influenced by both the Tx--Rx distance and the surrounding building geometry. In the denser scene, the propagation pattern is more fragmented due to frequent obstruction and multipath interactions, whereas a more open scene exhibits a clearer distance-dependent attenuation pattern with more extended low loss regions. 

\subsubsection{Impact of Antenna Tilt Angle}

In this work, the tilt angle $\alpha$ is defined as the angle between the antenna main beam direction and the horizontal plane, where a positive value represents a downward tilt. This parameter plays an important role in practical cellular systems, as it affects both the spatial distribution of received power and the trade-off between coverage and inter-cell interference \cite{Seifi2012}. 

For a given BS height $h_{\mathrm{BS}}$ and receiver height $h_{\mathrm{R}}$, the distance from the BS to the point where the main beam reaches the receiver plane is given by \cite{Thiele2009}:

\begin{equation}
d_{\mathrm{tilt}}
=
\frac{h_{\mathrm{BS}}-h_{\mathrm{R}}}{\tan\alpha},
\end{equation}

To evaluate the impact of the tilt angle, we replace the dipole antenna used in the preceding simulations with the directional 3GPP TR~38.901 antenna pattern. The directional pattern is adopted here because changes in antenna orientation can be more clearly reflected in the spatial distribution of antenna gain than with the previously used dipole pattern. The default half-power beamwidth (HPBW) of $65^\circ$ is used, while the tilt angle is selected with reference to \cite{Seifi2012}.

\begin{figure}[htbp]
    \centering
    \includegraphics[width=\columnwidth]{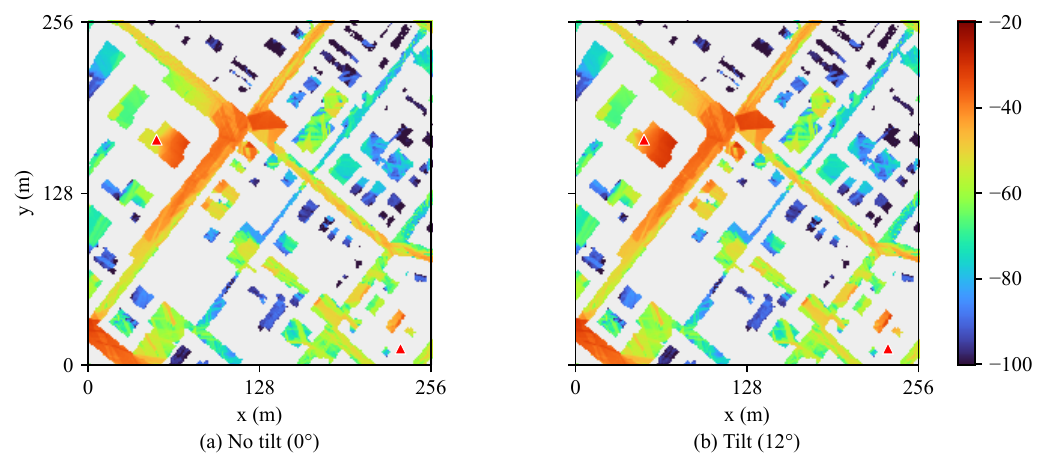}
    \footnotesize
    \caption{Results obtained with the directional 3GPP TR~38.901 antenna pattern under (a) no tilt ($0^\circ$) and (b) a $12^\circ$ downtilt.}
    \label{fig:tilt}
\end{figure}

As shown in Fig.~\ref{fig:tilt}, the overall propagation structures and coverage regions remain highly similar under the two tilt configurations, although moderate local variations in received power can be observed. This relatively small difference is mainly attributed to the large Half-Power Beamwidth (HPBW) adopted in the experiment. With a broad antenna beam, a $12^\circ$ rotation still leaves most receiver locations within or close to the main radiation region, resulting in only limited redistribution of antenna gain over the considered area.

\section{Conclusion}
This paper presented UrbanEMF, a large-scale EMF mapping dataset generated from a spatially continuous area of the Paris urban environment and real-world base-station deployments. The dataset provides both RSS and path loss maps while preserving realistic urban geometry, multi-BS information, and receiver height information. The resulting data can therefore support a wide range of radio-environment studies, from area level coverage and exposure analysis to transmitter specific propagation and link-level investigation.

Future work will extend the dataset pipeline to cover additional cities, carrier frequencies, and deployment conditions. This expansion will broaden the urban environments of the dataset. Temporal wireless network variations will also be investigated more thoroughly to capture dynamic real-world behaviors. In addition, richer channel-related outputs will be incorporated to increase the depth and representativeness of the data. which aims to improve both the diversity and the applicability of the dataset.

\bibliographystyle{IEEEtran}
\bibliography{references}

\end{document}